\documentclass[aps,12pt,preprintnumbers,nofootinbib,superscriptaddress]{revtex4}
\usepackage{graphicx}
\usepackage{amssymb,amsmath}
\usepackage{slashed}
\usepackage{dcolumn}%\usepackage{epstopdf}
\begin{document}

\title{\Large{\bf Aspects of a Nonminimal Conformal Extension of the Standard Model} }

\author{Feng Wu}
\email[Electronic address: ]{fengwu@ncu.edu.cn}
\affiliation{%
Department of Physics, Nanchang University, Nanchang
330031, China}

\date{\today}

\begin{abstract}
In this article we investigate a conformal extension of the standard model in which the scalar sector consists of a standard model Higgs doublet, a real gauge singlet and a real $SU(2)_{L}$ triplet. Focusing on the scenario where the Higgs boson found at the LHC is identified as the pseudo-Nambu-Goldstone boson of broken scale invariance, various theoretical and phenomenological features of the model are discussed. In particular, we analyze the decay pattern of the new scalar resonance. We also show that when the mass of this new scalar resonance is far below the $WW$ threshold, the natural regions of the parameter space are reduced by a constraint associated with the symmetry enhancement due to the decoupling of the singlet scalar from the rest of the system.  
\end{abstract}
\maketitle 
\newpage

\date{\today}

\section{Introduction}
The gauge hierarchy problem is one of the most serious conundrums in particle physics. In the standard model (SM) the electroweak symmetry breaking is triggered by the negative mass-squared term in the scalar potential. However, radiative corrections to the self-energy of the Higgs boson in the SM are quadratically sensitive to the new physics scale. If the new physics scale is significantly above the electroweak scale, which seems to be a possibility hinted by the current LHC data, an unnatural fine-tuning is required to render the electroweak scale stable. So far experimental searches at the LHC have shown no signs of supersymmetry or extra dimensions \cite{PDG}, which are ideas proposed by theorists for a natural solution to the gauge hierarchy problem. The stability of the electroweak scale thus remains a mystery to date.  

An alternative solution to the gauge hierarchy problem was proposed by Bardeen \cite{Bardeen}. Noticing that the SM Lagrangian without the Higgs mass term is conformally invariant at the classical level, Bardeen argued that once the classical conformal invariance is imposed on the SM, the quadratic divergences appearing in the quantum corrections to the Higgs mass can be removed at the UV cutoff scale as a boundary condition of the underlying theory in which the scale invariance would protect the Higgs mass from large radiative corrections, and so physical observables will no longer depend on them. In this approach, since there is no dimensionful parameter in the original Lagrangian, electroweak symmetry breaking is dynamically induced via the Coleman-Weinberg mechanism \cite{Coleman} and physical scales are generated by dimensional transmutation. However, within the SM particle content, the Coleman-Weinberg mechanism does not work due to the large negative contribution from the top-quark loop that keeps the effective potential from developing a minimum. Thus the extension of the bosonic sector is necessary to realize Bardeen's idea. 

Following this line of thought, many models along this direction have been constructed and various related issues have been studied \cite{1,2,3,4,5,6,7,8,9,10,101,11,12,13,14,15,16,17,18,19,20,21,22,23,24}. A recent analysis in \cite{Helmboldt} has shown that without changing the gauge sector of the SM, the minimal conformal extension of the SM that enables radiative symmetry breaking and generates no Landau poles or vacuum instabilities below the Planck scale in the renormalization-group evolution needs two real scale gauge singlets to enlarge the scalar sector of the SM, and at least one of them should develop a nonzero vacuum expectation value.

In this paper, we investigate a modified version of the conformal SM in which the enlarged scalar sector consists of a SM Higgs doublet, a real gauge singlet and a real $SU(2)_{L}$ triplet. In \cite{Helmboldt}, it has already been pointed out that replacing a real gauge singlet by a higher-dimensional real $SU(2)_{L}$ multiplet will cause possible Landau poles to appear at higher scales, and hence we will not analyze the stability of the running coupling constants any further. Instead, we restrict our attention on other theoretical and phenomenological consequences of the model. In Sec. II, after introducing the model we want to investigate, we work out the scalar mass spectrum after spontaneous symmetry breaking via the Coleman-Weinberg mechanism and discuss several distinct features of the model. In particular, we focus on the scenario where the Higgs boson discovered at the LHC is identified as the pseudo-Nambu-Goldstone boson of broken scale invariance. We then proceed to study the decay pattern of the new scalar degrees of freedom in the theory. In Sec. III, we consider the situation where two hierarchically separated scales are generated in the broken-symmetry phase and clarify that scale symmetry alone cannot protect the stability of the light scale. This fact would further constrain the parameter space of the model. Our conclusions are given in Sec. IV. In Appendix A, we analyze the case where all neutral components of the scalars in the theory have nonzero vacuum expectation values.  

\section{The Model}
Assume that the gauge group of the SM remains unchanged. The model we will investigate in this work is the extension of the SM in which the scalar sector is composed of the usual complex Higgs doublet $\Phi=(\phi^{+}\,\,\, \phi^{0})^{T}$ that transforms as $(1,2,{1\over2})$ under the gauge group $SU(3)_{C}\times SU(2)_{L}\times U(1)_{Y}$, a real triplet $\Sigma$ transforming as $(1,3,0)$ and a real singlet $\chi$ transforming as $(1,1,0)$. The real triplet $\Sigma$ can be expressed as
 \begin{equation}
\Sigma={1\over 2} \left( \begin{array}{cc}
\Sigma^{0}     & \sqrt{2} \Sigma^{+} \\
 \sqrt{2} \Sigma^{-}  & - \Sigma^{0}      \\
 \end{array} \right).
\end{equation} 
By imposing scale invariance$\footnote{It is believed that a unitary Poincare-invariant interacting field theory that is scale-invariant is also conformally invariant. \cite{Luty,Dymarsky,Dymarsky2} }$, the Lagrangian of the scalar sector is
\begin{equation}
\mathcal{L}_s=\left(D_{\mu}\Phi \right)^{\dagger} \left(D^{\mu}\Phi \right)+Tr \left(D_{\mu}\Sigma \right)^{\dagger} \left(D^{\mu}\Sigma \right)+{1\over 2} \partial_{\mu} \chi \partial^{\mu} \chi -V(\Phi,\Sigma,\chi) \label{Ls},
\end{equation}
where the covariant derivative of $\Phi$ and that of $\Sigma$ are given, respectively, by
\begin{equation}
D_{\mu} \Phi=\left( \partial_{\mu} -i g_2  W_{\mu}^{a} T^{a} -i {g_{1}\over 2} X_{\mu} \right) \Phi
\end{equation} 
and 
\begin{equation}
D_{\mu} \Sigma= \partial_{\mu}\Sigma -i g_{2}  [ W_{\mu}^{a} T^{a} , \Sigma].
\end{equation}
Here $W_{\mu}^{a},a=1,2,3$, and $X_{\mu}$ are the gauge bosons for the gauge groups $SU(2)_{L}$ and $U(1)_{Y}$, respectively, $g_{2}$ and $g_{1}$ are the corresponding gauge coupling constants, and $T^{a}$ are the generators of the gauge group $SU(2)_{L}$. The scalar potential $V(\Phi,\Sigma,\chi)$ reads
\begin{equation}
V(\Phi,\Sigma,\chi)=\lambda_{1} (\Phi^{\dagger} \Phi)^2+\lambda_{2}\chi^4 + \lambda_{3}Tr \Sigma^4 + \lambda_{4}(\Phi^{\dagger}\Phi)\chi^{2} + \lambda_{5}(\Phi^{\dagger} \Phi) Tr \Sigma^2 +\lambda_{6} \chi^{2} Tr \Sigma^2 +g \chi \Phi^{\dagger} \Sigma \Phi, \label{V}
\end{equation}
where all coupling constants are dimensionless. Note that $(Tr \Sigma^2)^2=2Tr \Sigma^4$ and $\Phi^{\dagger} \Sigma^2 \Phi={1\over 2} (\Phi^{\dagger} \Phi)Tr \Sigma^2$. The potential~(\ref{V}) is thus the most general renormalizable one. In addition, it is useful to know that without the last term, the above scalar potential possesses an accidental global symmetry $O(4)_{\Phi}\times O(3)_{\Sigma}$. Hence a small value of $g$ is natural.  

In general, the electrically neutral component of each scalar can radiatively develop a nonzero vacuum expectation value, breaking the conformal invariance and possibly part of the electroweak gauge symmetry. Let us parametrize the neutral scalar fields as $  \phi^{0}=v_{\phi} +{1\over \sqrt{2}} (h+i \delta)$, $\chi=v_{\chi}+ \eta$ and $\Sigma^{0}=v_{\Sigma}+\sigma$, with $v_{i}$ being the vacuum expectation value of the scalar $i$. In order to be consistent with phenomenology, we know that $v_{\phi}\neq 0$ is required. Also, the possibility that neither $\chi$ nor $\Sigma^{0}$ acquires a nonzero vacuum expectation value is ruled out due to the appearing of Landau poles below the Planck scale \cite{Helmboldt}. Thus, we are left with the following two cases$\footnote{As shown in Appendix A, the case in which $v_{\phi}v_{\chi}v_{\Sigma} \neq 0$ leads to a tree-level correction to the $\rho$-parameter: 
\begin{equation}
\rho=1+2\left( {v_{\Sigma}\over v_{\phi}} \right)^2. \label{rho}
\end{equation}
 From the global fit to electroweak precision data, this case is not phenomenologically favorable compared to the one with $v_{\Sigma}=0$. }$:
\[
{\rm(i)}\,\, v_{\phi} v_{\chi} \neq 0 \,\,\,{\rm{and}}\,\,\, v_{\Sigma} =0;\,\,\,\,\,\,\, {\rm(ii)} \,\, v_{\phi} v_{\Sigma} \neq 0 \,\,\,{\rm{and}}\,\,\, v_{\chi} =0.
\] 
Before we proceed, let us note that either case (i) or case (ii) is possible only when $g=0$ at the symmetry breaking scale. This fact can be easily seen from the minimization conditions ${\partial V \over \partial v_{\Sigma}}=0$ and ${\partial V \over \partial v_{\Sigma}}=0$, respectively, for each case. In the case where $v_{\Sigma}=0$, without symmetry breaking at other scales, the $O(3)_{\Sigma}$ symmetry will keep the value of the coupling constant $g$ vanishing at all scales. Similarly, in the case where $v_{\chi}=0$, the value of $g$ remains zero at all scales due to the symmetry under $\chi\longrightarrow -\chi$ unless the field $\chi$ develops a nonzero vacuum expectation value at some other scale.

For the case (ii), it is easy to show that there are four massless charged scalars. Two of them are the would-be Nambu-Goldstone bosons. The other two massless charged scalars $\Sigma^{\pm}$ are the Nambu-Goldstone bosons associated with the breaking of the global $O(3)_{\Sigma}$ symmetry. From the kinetic term $Tr (D_{\mu} \Sigma)^{\dagger} (D^{\mu} \Sigma)$, one can further show that the couplings of the scalars $\Sigma^{\pm}$ to the gauge bosons $W_{\mu}^{a}$ are not suppressed. Thus, apart from affecting the $\rho$-parameter, this feature also makes the case (ii) phenomenologically unviable. In what follows, we will concentrate on the case (i) and neglect the last term in~(\ref{V}).

\subsection{mass spectrum} 
Particles acquire their masses through spontaneous symmetry breaking. Instead of calculating explicitly the full one-loop effective potential and then minimizing it, another way to obtain the nonzero vacuum expectation values is by using the elegant method developed by Gildener and Weinberg \cite{Gildener}. In this approach, one first uses the minimization conditions for the tree-level potential to find certain relations between coupling constants at the symmetry breaking scale, denoted by $\Lambda_{GW}$. These relations specify a line of degenerate minima passing through the origin of field space. Then the one-loop effective potential along this flat direction will determine the real vacuum state. Adopting the notation of Ref.\cite{Helmboldt} by writing
\begin{eqnarray}
v_{\phi}=\sin \alpha \langle \phi \rangle,\\
v_{\chi}=\cos \alpha \langle \phi \rangle,  
\end{eqnarray} 
it is straightforward to show that a flat direction exists when $\alpha$ is defined by the following relation at the scale $\Lambda_{GW}$:
\begin{equation}
\tan^{2}\alpha=-{\lambda_{4} \over 2 \lambda_{1}}= -{2\lambda_{2} \over  \lambda_{4}}. \label{alpha}
\end{equation}
This relation tells us that $\lambda_{1}\lambda_{2} > 0$ and $\lambda_{1,2}\lambda_{4} < 0$. Here, we should note that the above relation holds only at $\Lambda_{GW}$. The dominant running behavior of the coupling constants $\lambda_{1}$, $\lambda_{2}$, and $\lambda_{4}$ is governed by the one-loop $\beta$ functions, given by
\begin{eqnarray}
\beta_{\lambda_{1}}= {d \lambda_{1}(\mu)\over d\ln \mu}&=&{1\over 16 \pi^2} \left(24\lambda_{1}^2 + 2 \lambda_{4}^2+{3\over 2} \lambda_{5}^{2} + \left( -3 \lambda_{1}(g_{1}^2+3g_{2}^2)+{3\over 8} (g_{1}^{4} +3 g_{2}^{4} + 2g_{1}^{2}g_{2}^{2})\right) \right.\\
  & &\left. +6(2\lambda_{1}-y_{t}^2)y_{t}^2\right), \label{beta1}\\
\beta_{\lambda_{2}}= {d \lambda_{2}(\mu)\over d\ln \mu}&=& {1\over 16 \pi^2}\left(72\lambda_{2}^{2}+2\lambda_{4}^{2}+{3\over 2} \lambda_{6}^{2}\right), \label{beta2}\\
 \beta_{\lambda_{4}}= {d \lambda_{4}(\mu)\over d\ln \mu}&= &{1\over 16 \pi^2}\left( 8\lambda_{4}({3\over 2}\lambda_{1}+3\lambda_{2}+\lambda_{4}) +3\lambda_{5}\lambda_{6} -{3\over 2}\lambda_{4}(g_{1}^{2}+3g_{2}^{2})+6\lambda_{4} y_{t}^2 \right),\label{beta3}
\end{eqnarray}
where $y_{t}$ is the top-quark Yukawa coupling constant.

The tree-level masses of the scalars can be obtained from the scalar potential~(\ref{V}) with $g=0$. There are eight scalar degrees of freedom in the model. Among them, the massless fields $\delta$, $\phi{\pm}$ are would-be Goldstone bosons that will be eaten, respectively, by the $Z^{0}$ and $W_{\pm}$ upon electroweak symmetry breaking. The squared masses of the scalars $\Sigma_{\pm} $ and $\Sigma_{0}$ are degenerate, and given by
\begin{equation}
m_{\Sigma^{\pm}}^{2}=m_{\Sigma^{0}}^{2}=\lambda_{5} v_{\phi}^{2} + \lambda_{6} v_{\chi}^{2} .\label{sigmamass}
\end{equation}
The unbroken $O(3)_{\Sigma}$ symmetry insures that this degeneracy is preserved.

The tree-level mass terms for the $CP$-even neutral scalars $(h, \eta)$ are given by
\begin{equation}
V(\Phi,\Sigma,\chi)\supset {1\over 2} \left(h\,\,\, \eta\right) {\bf{M}}^2 \left( \begin{array}{c}
h  \\
 \eta \\
 \end{array} \right),
\end{equation}
where the mass matrix $\bf{M}^2$ at the scale $\Lambda_{GW}$ takes the form
\begin{equation}
{\bf{M}}^{2}= \left( \begin{array}{cc}
4\lambda_{1} v_{\phi}^2     & 2\sqrt{2} \lambda_{4} v_{\phi}v_{\chi} \\
 2\sqrt{2} \lambda_{4} v_{\phi}v_{\chi} & - 4\lambda_{4} v_{\phi}^2    
 \end{array} \right).\label{hmass}
\end{equation} 
The minimization conditions and the relation~(\ref{alpha}) have been used to obtain the above expression. One can see that the Higgs portal coupling constant $\lambda_{4}$ causes the mixing between $h$ and $\eta$. The eigenvalues of this mass matrix are the tree-level squared masses of physical scalars, denoted by $h_{1}$ and $h_{2}$. Define the mixing angle $\theta$ to be the angle that comes out in the change of basis from $(h, \eta)$ to the physical one $(h_{1},h_{2})$:
\begin{equation}
\left( \begin{array}{c}
h_{1} \\
h_{2}    
 \end{array} \right)= \left( \begin{array}{cc}
\cos \theta & -\sin\theta \\
\sin\theta & \cos\theta     
 \end{array} \right) \left( \begin{array}{c}
h \\
\eta    
 \end{array} \right) .\label{theta}
\end{equation} 
Without loss of generality, we choose $\vert \tan\theta\vert <1$, so that $h_{1}$ is mainly composed of the doubletlike neutral scalar $h$, and hence is identified as the Higgs boson discovered at the LHC. It is then easy to show that
\begin{equation}
\tan\theta= \left\{ \begin{array}{c}
\sqrt{2} \tan\alpha \,\,\,\,\, {\rm{for}} \,\,\,\theta >0, \\
 -{1\over \sqrt{2} \tan\alpha }\,\,\,\,\, {\rm{for}} \,\,\,\theta <0.   
 \end{array} \right. \label{hmass}
\end{equation}  
Assume that $v_{\chi}<v_{\phi}$, so that $\theta <0$. With this choice, we find that the dynamically generated masses along the flat direction are given by  
\begin{eqnarray}
m_{h_{1}}^{2}&=&0,\\
m_{h_{2}}^{2}&=&4(\lambda_{1}-\lambda_{4})v_{\phi}^{2}\label{h2mass}.  
\end{eqnarray} 
Note that the physical scalar $h_{1}$ is the pseudo-Nambu-Goldstone boson of broken scale invariance, and is massless only at tree-level$\footnote{Recently, a detailed phenomenological study of dilaton interactions is presented in \cite{Bandyopadhyay}, where current data are used to analyze the signatures and the bounds on a possible dilaton state at the LHC. However, we should note that different from the scenario in our work, the LHC Higgs boson is not identified as the dilaton there.}$. The radiative mass for $h_{1}$ induced by the one-loop effective potential, when identified with the Higgs boson mass measured at the LHC, would provide a constraint on the parameters of the model$\footnote{The current experimental bound on the mixing angle $\theta$ is $\vert \tan \theta \vert < 0.4$ \cite{Farzinnia}. This tells us that if the Higgs boson found at the LHC does not correspond to the pseudo-Nambu-Goldstone boson, we must find $v_{\chi}>v_{\phi} $.}$. Since the result is not illuminating, we will not include it here. 

Before we proceed, let us remark some features of the model. First note that the Yukawa Lagrangian for the fermion sector is the same as the one in the SM. Thus, similar to the gauge boson masses, the tree-level fermion masses depend only on the vacuum expectation value $v_{\phi}$, whereas we have seen that the tree-level masses of the scalars might depend also on $v_{\chi}$. Secondly, as shown clearly in~(\ref{rho}), the $\rho$-parameter does not deviate from 1 at tree-level for $v_{\Sigma}=0$. This is a phenomenologically encouraging fact. Thirdly, due to the exact $O(3)_{\Sigma}$ symmetry, there are no light particles into which $\Sigma^{0}$ and $\Sigma^{\pm}$ can decay. The stable electrically neutral particle $\Sigma^{0}$ thus makes an attrictive candidate for the dark matter \cite{Perez}.

\subsection{$h_{2}$ decay}   
We now investigate the decay modes of the neutral scalar $h_{2}$. The decay pattern depends on the values of parameters. First, note that for the scalars $h_{1}$, $h_{2}$, $\Sigma^{0}$, $\Sigma^{\pm}$ to be physical particles, the relation~(\ref{alpha}) and the mass formulae~(\ref{sigmamass}) and~(\ref{h2mass}) tell us that the following inequalities should be satisfied at the symmetry-breaking scale:
\begin{equation}
\lambda_{1}>0,\,\,\, \lambda_{2}>0, \,\,\, \lambda_{4} <0, \,\,\, \lambda_{5}\lambda_{4}<2\lambda_{1}\lambda_{6}. \label{condition}
\end{equation}  
The last inequality implied that $\lambda_{5}$ and $\lambda_{6}$ cannot both be negative. 

In general, the decay branching ratios for $h_{2}$ are different from those of the SM Higgs boson ($h_{1}$). Exploring the decay phenomenology of $h_{2}$, if existing, is crucial to the reconstruction of the scalar potential, which in turn will give us further insights into the precise realization of the Higgs mechanism.

When ${1\over 2}m_{h_{1}} < m_{h_{2}} <2 m_{W}$, the $h_{2}$ and $h_{1}$ branching ratios will be identical, with $h_{2}$ being a narrower resonance provided that the condition $v_{\chi} <v_{\phi}$ is met. When $2m_{Z}< m_{h_{2}} <2 m_{h_{1}}$, other than the same decay modes as those of the SM Higgs boson, decay channels $h_{2}\rightarrow W^{+}W^{-},\,\,ZZ$ are permitted, with the tree-level partial widths given by
\begin{equation}
\Gamma_{h_{2}\rightarrow W^{+}W^{-}}={1\over32\pi} {m_{h_{2}}^{2}\over v_{\phi}^{2}}\sin^{2} \theta \left( 1-4{m_{W}^{2}\over m_{h_{2}}^{2}} +12 {m_{W}^{4} \over m_{h_{2}}^{4}} \right)\left(  m_{h_{2}}^{2}- 4m_{W}^{2}  \right)^{1/2},
\end{equation}
and
\begin{equation}
\Gamma_{h_{2}\rightarrow Z Z}={1\over 2}\Gamma_{h_{2}\rightarrow W^{+}W^{-}}(m_{W}\rightarrow m_{Z}).
\end{equation}
We note in passing that in this model $hW^{+}W^{-}$ and $hZZ$ couplings are the only cubic interactions between two gauge bosons and one scalar at tree-level. In the case where the mass of $h_{2}$ is above the $h_{1}h_{1}$ threshold, the Higgs splitting mode $h_{2}\rightarrow h_{1}h_{1}$ is also allowed. It is straightforward to work out from the cubic terms of the scalar potential~(\ref{V}), rewriting in the basis of mass eigenstates $(h_{1},h_{2})$, that
\begin{equation}
V\supset 3\sqrt{2} v_{\phi} \sin\theta \left( \lambda_{1} \cos^{2}\theta -4\lambda_{2} \sin^{2}\theta - \lambda_{4}\cos 2\theta    \right) h_{2} h_{1}h_{1},
\end{equation}   
and hence the tree-level partial width of the Higgs splitting mode is given by
\begin{equation}
\Gamma_{h_{2}\rightarrow h_{1}h_{1}} = {9\over 4\pi}{v_{\phi}^{2}\over m_{h_{2}}^{2}}\sin^{2}\theta  [ \lambda_{1} \cos^{2}\theta -4\lambda_{2} \sin^{2}\theta - \lambda_{4}\cos 2\theta   ]^2 \left(  m_{h_{2}}^{2}- 4m_{h_{1}}^{2}  \right)^{1/2}.
\end{equation}
Notice that the quantity in brackets vanishes at the symmetry-breaking scale $\Lambda_{GW}$. This can be easily checked by using~(\ref{alpha}). Its value at any other scale is given by the solutions to the renormalization group equations~(\ref{beta1}), (\ref{beta2}) and (\ref{beta3}).

A modification of the above discussion is required when the following inequality among the parameters is satisfied:
\begin{equation}
\lambda_{1}(\lambda_{4}+2\lambda_{6}) < \lambda_{4}( \lambda_{4} + \lambda_{5}). \label{condition2}
\end{equation}
In this circumstance, $m_{h_{2}} > 2 m_{\Sigma}$ is guaranteed. Thus, except decaying into the SM particles, $h_{2}$ can also decay into $\Sigma$ particles. Again, from the scalar potential we easily find
\begin{equation}
V\supset \sqrt{2} v_{\phi} \sin\theta \left({\lambda_{5}\over 2}-\lambda_{6}\right) h_{2}(\Sigma^{0}\Sigma^{0} +2 \Sigma^{+}\Sigma^{-}),
\end{equation}
and it follows that
\begin{equation}
\Gamma_{h_{2}\rightarrow \Sigma^{0}\Sigma^{0}}={1\over 2}\Gamma_{h_{2}\rightarrow \Sigma^{+}\Sigma^{-}}={1\over4\pi}{v_{\phi}^{2}\over m_{h_{2}}^{2}} \sin^{2}\theta \left({\lambda_{5}\over 2}-\lambda_{6}\right)^{2} \left(  m_{h_{2}}^{2}- 4m_{\Sigma}^{2}  \right)^{1/2}.
\end{equation}

\section{Naturalness}
A theory possessing a hierarchical structure is usually unnatural. The primary motivation for exploring the extension of the SM with conformal invariance is to avoid the gauge hierarchy problem without any unnatural fine-tuning. However, when the absolute value of the mixing angle $\theta$ is very small, that is, when $\lambda_{1} \ll -\lambda_{4} \ll \lambda_{2}$, the relation~(\ref{alpha}) indicates that the model contains two hierarchically separated scales $v_{\chi} \ll v_{\phi}$. In this case, the fine-tuning issue reappears. 

To be more explicit, let us assume that the coupling constants $\lambda_{2}$, $\lambda_{5}$, $\lambda_{6}$ are all of order 1. Then from the tree-level mass formulae~(\ref{sigmamass}) and (\ref{h2mass}), we know that the masses of the scalars $\Sigma^{0}$ and $\Sigma^{\pm}$ are of order $v_{\phi}$, while the dynamically induced mass of the pseudo-Nambu-Goldstone boson $h_{1}$ is suppressed by a loop factor compared to the $\Sigma $ mass. On the other hand, using the relation~(\ref{alpha}), we find that the tree-level mass of the scalar $h_{2}$ is of order $v_{\chi}$, which is very light compared to the masses of other bosons in the model. This corresponds to the situation in which the mass of $\phi_{2}$ is far below the $WW$ threshold and is an extremely narrow resonance$\footnote{When the condition $m_{h_{2}} < {1\over 2} m_{h_{1}}$ is satisfied, the model predicts a nonstandard decay mode $h_{1}\rightarrow h_{2}h_{2}$ for the observed Higgs boson, with the partial width being given by
\begin{equation}
\Gamma_{h_{1}\rightarrow h_{2}h_{2}}\simeq {\lambda_{4}^{2}v_{\phi}^{2}\over \pi m_{h_{1}}^{2}}(m_{h_{1}}^{2} + 16 \lambda_{4} v_{\phi}^{2} )^{1/2}.
\end{equation}   }$.

Now, let us consider the quantum effects. It is not difficult to verify that due to the coupling of the singlet scalars $\chi$ to the heavy scalars $\Sigma$, the radiative correction receiving from the $\Sigma$ loop to the mass-squared for $h_{2}$, denoted by $\delta m_{h_{2}}^{2}$, is$ \footnote{ When $v_{\chi} \ll v_{\phi}$, the neutral scalar $h_{2}$ is composed almost entirely of the singlet scalar $\eta$. }$
\begin{equation}
\delta m_{h_{2}}^{2} \sim \lambda_{6} v_{\phi}^{2} \gg v_{\chi}^{2}
\end{equation}
for $\lambda_{6} \sim O(1)$. Thus, the presence of the light physical scale is unnatural, unless the values of $\lambda_{6}$ and $\lambda_{4}$ are of the same order, that is, 
\begin{equation}
\vert \lambda_{6} \vert \sim \vert \lambda_{4} \vert \ll 1.
\end{equation}
Once the above restriction on $\lambda_{6}$ is satisfied, the light physical mass will no longer be sensitive to the heavy scale in the theory, and is technically natural. From this analysis we conclude that naturalness imposes an additional constraint on the parameters of the the model.

Notice that in the limit of vanishing $\lambda_{4}$ and $\lambda_{6}$, the $\chi$ sector decouples from the rest of the system. In this limit, the theory enjoys an extended Poincare symmetry, meaning that performing independent Poincare transformations on the $\chi$ sector and the rest of the system leaves each part invariant individually. As was emphasized in Ref.\cite{Foot}, scale invariance alone does not necessarily secure the stability of the light scale and in this case it is the emerging extended Poincare symmetry that protects the light scale from large radiative corrections and guarantees its naturalness.

\section{Conclusion}
The fact that no new physics has been observed so far at the LHC machine  drives us to sharpen our understanding of the stability of electroweak symmetry-breaking scale. In this paper, we have investigated a specific extension of the conformal SM in which masses of particles are generated dynamically through the dimensional transmutation, presumably stable against the corrections. Keeping in mind that the model can remain stable under renormalization group translations below the Planck scale and restricting our attention to the scalar sector, we have discussed several characteristic features of the model. Also, we have analyzed the decay pattern of new scalar resonances, which is controlled by the parameters of the theory.

On the other hand, although the primary motivation for this approach is to avoid the gauge hierarchy problem, we have demonstrated that when the mixing between the SM-like and the singlet-like scalars in the model is negligible, the stability of the hierarchical structure is actually protected not by scale symmetry, but by the enhanced spacetime symmetry in the decoupling limit described in Sec. III. The data of precision Higgs boson measurements and the outcome of experimental search for additional scalars in the near future will definitely give further insights that will help us realize if this approach is accurate.

\begin{acknowledgments}
This research was supported in part by the National Nature Science Foundation of China under Grant No.11565019 and the 555 talent project of Jiangxi Province.
\end{acknowledgments}

\appendix
\section{The case $v_{\phi}v_{\chi}v_{\Sigma}\neq 0$}
Here we consider the case where every neutral component of the scalars in~(\ref{V}) develops a nonzero vacuum expectation value from the spontaneous symmetry breaking. It is straightforward to obtain the tree-level gauge boson masses from the kinetic energy terms of $\Phi$ and $\Sigma$. We find
\begin{equation}
m_{W}^{2}=g_{2}^{2}\left( {1\over 2} v_{\phi}^{2} + v_{\Sigma}^{2} \right),\,\,\,\,\, m_{Z}^{2}={1\over 2}v_{\phi}^2\left(g_{2}^{2}+g_{1}^{2}  \right),
\end{equation} 
and hence the tree-level $\rho$-parameter is 
\begin{equation}
\rho \equiv {m_{W}^{2} \over m_{Z}^{2} \cos^2 \theta_{W}}=1+2\left({v_{\Sigma}\over v_{\phi} }\right)^2,\label{rho1}
\end{equation} 
where $\theta_{W}\equiv \tan^{-1}({g_1 \over g_2})$ is the usual weak mixing angle. From the global fit \cite{PDG}, we obtain an upper bound on the ratio of vacuum expectation values:
\begin{equation}
{v_{\Sigma}\over v_{\phi}} < 0.02.
\end{equation}
 
The mass terms for the charged scalars $(\Sigma^{\pm}, \phi^{\pm})$ come from the expansion of the potential~(\ref{V}) about the vacuum states. After some algebra, we obtain
\begin{equation}
V(\Phi,\Sigma,\chi)\supset  \left(\Sigma^{-}\,\,\, \phi^{-} \right) \left( \begin{array}{cc}
{1\over 2} g {v_{\chi} v_{\phi}^{2} \over v_{\Sigma}}    & {1\over \sqrt{2}} g v_{\chi} v_{\phi} \\
 {1\over \sqrt{2}} g v_{\chi} v_{\phi} & g   v_{\chi} v_{\Sigma}  
 \end{array} \right)\left( \begin{array}{c}
\Sigma^{+}  \\
 \phi^{+} \\
 \end{array} \right),\label{chargedmass}
\end{equation}
where the minimization conditions for the potential has been used to obtain the above expression. The eigenvalues of the above tree-level mass matrix are 0 and ${1\over 2} \rho g {v_{\chi}v_{\phi}^{2}\over v_{\Sigma}}$, with $\rho$ given in~(\ref{rho1}). Notice that when $g=0$, all eigenstates are massless. In this case, besides the two would-be Nambu-Goldstone bosons $\phi^{\pm}$, the other two massless eigenstates $\Sigma^{\pm}$ are the Nambu-Goldstone bosons arising through the spontaneous breaking of the global $O(3)_{\Sigma}$ symmetry.

Similarly, the mass terms for the $CP$-even neutral scalars in the basis $(h, \eta, \sigma)$ are given by 
\begin{equation}
V(\Phi,\Sigma,\chi)\supset {1\over 2} \left(h\,\,\, \eta\,\,\,\sigma \right) {\tilde{\bf{M}}}^2 \left( \begin{array}{c}
h  \\
 \eta \\
\sigma
 \end{array} \right),
\end{equation}
where the symmetric mass matrix $\tilde{\bf{M}}^2$ takes the form
\begin{equation}
{\tilde{\bf{M}}}^{2}= \left( \begin{array}{ccc}
4\lambda_{1} v_{\phi}^2     & -\sqrt{2} {v_{\chi}\over v_{\phi}}(4\lambda_{2}v_{\chi}^{2}+\lambda_{6}v_{\Sigma}^{2})& -\sqrt{2} {v_{\Sigma}\over v_{\phi}}({\lambda_{3}\over 2}v_{\Sigma}^{2}+\lambda_{6}v_{\chi}^{2})\\
 -\sqrt{2} {v_{\chi}\over v_{\phi}}(4\lambda_{2}v_{\chi}^{2}+\lambda_{6}v_{\Sigma}^{2}) & 8\lambda_{2} v_{\chi}^2+{1\over 2} g {v_{\Sigma}v_{\phi}^{2}\over v_{\chi}} & 2\lambda_{6}v_{\chi}v_{\Sigma}-{g\over 2} v_{\phi}^{2} \\
-\sqrt{2} {v_{\Sigma}\over v_{\phi}}({\lambda_{3}\over 2}v_{\Sigma}^{2}+\lambda_{6}v_{\chi}^{2})& 2\lambda_{6}v_{\chi}v_{\Sigma}-{g\over 2} v_{\phi}^{2}   &  \lambda_{3} v_{\Sigma}^{2} +{g\over 2}{v_{\chi}v_{\phi}^{2}\over v_{\Sigma}}
 \end{array} \right).
\end{equation} 
Again, we have used the minimization conditions on the potential to obtain the above result. It is easy to check that the determinant of ${\tilde{\bf{M}}}^{2}$ vanishes, meaning that at least one of the mass eigenvalues is zero. To be more explicit, the first row plus the second row multiplied by ${v_{\chi}\over \sqrt{2} v_{\phi}}$ plus the third row multiplied by ${v_{\Sigma}\over \sqrt{2} v_{\phi}}$ gives a row that is all zeros. This is exactly what we would expect from the fact that a massless field arises through spontaneous symmetry breaking of scale invariance. The other two eigenvalues of the mass matrix ${\tilde{\bf{M}}}^{2}$ are given by 
\begin{equation}
m_{\pm}^{2}={1\over 2} \left( Tr{\tilde{\bf{M}}}^{2}\pm\left((Tr{\tilde{\bf{M}}}^{2})^2 -4 \displaystyle\sum_{i=1}^{3} m_{ii} \right)^{1/2} \right),
\end{equation}
where $m_{ii}$ is the minor of the diagonal element $({\tilde{\bf{M}}}^{2})_{ii}$.

We would like to emphasize that without $O(3)_{\Sigma}$ symmetry, the phenomenology of the scalars in the model can be drastically different from the one in the case $v_{\Sigma}=0$. For example, in the situation when $v_{\Sigma}\neq 0$, the masses of the charged and neutral components of the triplet $\Sigma$ are no longer degenerate. Also, all three neutral physical scalars can generally decay into the light SM particles, and we no longer have a dark matter candidate in the model.

%-----------------------------------------------------------------------------------------------------------------------

%-----------------------------------------------------------------------------------------------------------------------


\begin{thebibliography}{99}

 \bibitem {PDG}K. A. Olive et al. (Particle Data Group), Chin. Phys. C\ \textbf{38}, 090001 (2014).

\bibitem {Bardeen}W. A. Bardeen, FERMILAB-CONF-95-391-T.

\bibitem {Coleman}S. R. Coleman and E. J. Weinberg, Phys. Rev. D\ \textbf{7}, 1888 (1973).

\bibitem {1}R. Hempfling, Phys. Lett. B\ \textbf{379}, 153 (1996).

\bibitem {2}K. A. Meissner and H. Nicolai, Phys. Lett. B\ \textbf{648}, 312 (2007).

\bibitem {3}R. Foot, A. Kobakhidze, and R. R. Volkas, Phys. Lett. B\ \textbf{655}, 156 (2007).

\bibitem {4}R. Foot, A. Kobakhidze, K. L. McDonald, and R. R. Volkas, Phys. Rev. D\ \textbf{76}, 075014 (2007).

\bibitem {5}T. Hambye and M. H. G. Tytgat, Phys. Lett. B\ \textbf{659}, 651 (2008).

\bibitem {6}R. Foot, A. Kobakhidze, and R. R. Volkas, Phys. Rev. D\ \textbf{84}, 075010 (2011).

\bibitem {7}C. Englert, J. Jaeckel, V. V. Khoze, and M. Spannowsky, JHEP \ \textbf{04}, 060 (2013).

\bibitem {8}V. V. Khoze and G. Ro, JHEP \ \textbf{10}, 075 (2013).

\bibitem {9}C. D. Carone and R. Ramos, Phys. Rev. D\ \textbf{88}, 055020 (2013).

\bibitem {10}C. T. Hill, Phys. Rev. D\ \textbf{89}, 073003 (2014).

\bibitem {101}E. Gabrielli, M. Heikinheimo, K. Kannike, A. Racioppi, M. Raidal, and C. Spethmann, Phys. Rev. D\ \textbf{89}, 015017 (2014).

\bibitem {11}A. Salvio and A. Strumia, JHEP \ \textbf{06}, 080 (2014).

\bibitem {12}A. Gorsky, A. Mironov, A. Morozov, and T. N. Tomaras, J. Exp. Theor. Phys. \ \textbf{120}, 344 (2015).

\bibitem {13}S. Benic and B. Radovcic, JHEP \ \textbf{01}, 143 (2015).

\bibitem {14}H. Okada and Y. Orikasa, Phys. Lett. B\ \textbf{760}, 558 (2016).

\bibitem {15}M. Lindner, S. Schmidt, and J. Smirnov, JHEP \ \textbf{10}, 177 (2014).

\bibitem {16}P. Humbert, M. Lindner, and J. Smirnov, JHEP \ \textbf{06}, 035 (2015).

\bibitem {17}C. D. Carone and R. Ramos, Phys. Lett. B\ \textbf{746}, 424 (2015).

\bibitem {18}P. Humbert, M. Lindner, S. Patra, and J. Smirnov, JHEP \ \textbf{09}, 064 (2015).

\bibitem {19}A. Latosinski, A. Lewandowski, K. A. Meissner, and H. Nicolai, JHEP \ \textbf{10}, 170 (2015).

\bibitem {20}A. Farzinnia, Phys. Rev. D\ \textbf{92}, 095012 (2015).

\bibitem {21}A. Ahriche, K. L. McDonald, and S. Nasri, JHEP \ \textbf{02}, 038 (2016).

\bibitem {22}A. Karam and K. Tamvakis, Phys. Rev. D\ \textbf{92}, 075010 (2015).

\bibitem {23}N. Haba, H. Ishida, N. Okada, and Y. Yamaguchi, Phys. Lett. B\ \textbf{754}, 349 (2016).

\bibitem {24}N. Haba, H. Ishida, R. Takahashi, and Y. Yamaguchi, JHEP \ \textbf{02}, 058 (2016).

\bibitem {Helmboldt}A. J. Helmboldt, P. Humbert, M. Lindner, and J. Smirnov, arXiv:1603.03603.

\bibitem {Luty}M. Luty, J. Polchinski and R. Rattazzi, JHEP\ \textbf{01}, 152 (2013).

\bibitem {Dymarsky}A. Dymarsky, Z. Komargodski, A. Schwimmer and S. Theisen, JHEP \ \textbf{10}, 171 (2015).

 \bibitem {Dymarsky2}A. Dymarsky, K. Farnsworth, Z. Komargodski, M. Luty and V. Prilepina, arXiv:1402.6322.
 
 \bibitem {Gildener}E. Gildener and S. Weinberg, Phys. Rev. D\ \textbf{13}, 3333 (1976).
 
 \bibitem {Perez}P. F. Perez, H. H. Patel, M. J. Ramsey-Musolf, and K. Wang, Phys. Rev. D\ \textbf{79}, 055024 (2009).

 \bibitem {Bandyopadhyay}P. Bandyopadhyay, C. Coriano, A. Costantini, and L. D. Rose, arXiv:1607.01933.
 
 \bibitem {Farzinnia}A. Farzinnia, H.-J. He, and J. Ren, Phys. Lett. B\ \textbf{727}, 141 (2013).
 
 \bibitem {Foot}R. Foot, A. Kobakhidze, K. L. McDonald, and R. R. Volkas, Phys. Rev. D\ \textbf{89}, 115018 (2014). 

 
 
\end{thebibliography}
\end{document}